\documentclass{aa}  

\usepackage{graphicx}
\usepackage{txfonts}
\usepackage[colorlinks=true,
linkcolor=blue,
citecolor=blue,
urlcolor=blue,
plainpages=true        
]{hyperref}     %

\usepackage{pifont}
\usepackage{ulem}

\newcommand{\id}{\mathrm{d}}

\newcommand{\Md}{M_{\rm d}}
\newcommand{\rin}{r_{\mathrm{in}}}
\newcommand{\rout}{r_{\mathrm{out}}}

\newcommand{\Msun}{M_{\odot}}

\newcommand{\bcheckmark}{\textcircled{\small{\ding{51}}}}

\begin{document} 
    
   \titlerunning{Dynamics and origin of TRAPPIST-1}

   \title{TRAPPIST-1: Dynamical analysis of the transit-timing variations and origin of the resonant chain\thanks{Table 1 is only available in electronic form at the CDS via anonymous ftp to cdsarc.u-strasbg.fr (130.79.128.5) or via http://cdsweb.u-strasbg.fr/cgi-bin/qcat?J/A+A/}}

   \subtitle{}

   \author{J. Teyssandier\inst{1}\fnmsep\thanks{E-mail: jean.teyssandier@unamur.be}, A-S. Libert\inst{1}
          \and
          E. Agol\inst{2}
          }

   \institute{naXys, Namur Institute for Complex Systems, Department of Mathematics, University of Namur, 61 Rue de Bruxelles, 5000 Namur, Belgium
         \and
             Astronomy Department and Virtual Planetary Laboratory, University of Washington, Seattle, WA 98195, USA
             }

   \date{Accepted XXX. Received YYY; in original form ZZZ}

 
  \abstract
   {We analyze solutions drawn from the recently published posterior distribution of the TRAPPIST-1 system, which consists of seven Earth-size planets appearing to be in a resonant chain around a red dwarf. We show that all the planets are simultaneously in two-planet and three-planet resonances, apart from the innermost pair for which the two-planet resonant angles circulate. By means of a frequency analysis, we highlight that the transit-timing variation (TTV) signals possess a series of common periods varying from days to decades, which are also present in the variations of the dynamical variables of the system. Shorter periods (e.g., the TTVs characteristic timescale of 1.3~yr) are associated with two-planet mean-motion resonances, while longer periods arise from three-planet resonances. By use of $N$-body simulations with migration forces, we explore the origin of the resonant chain of TRAPPIST-1 and find that for particular disc conditions, a chain of resonances -- similar to the observed one -- can be formed which accurately reproduces the observed TTVs. Our analysis suggests that while the 4-yr collected data of observations hold key information on the two-planet resonant dynamics, further monitoring of TRAPPIST-1 will soon provide signatures of three-body resonances, in particular the 3.3 and 5.1~yr periodicities expected for the current best-fit solution. dditional observations would help to assess whether the innermost pair of planets is indeed resonant (its proximity to the 8:5 resonance being challenging to explain), and therefore give additional constraints on formation scenarios.
   }


   \maketitle
%

\section{Introduction}
\label{sec:intro}

The TRAPPIST-1 planetary system consists of at least seven Earth-size planets around an ultra-cool red dwarf star of mass $M_*\approx 0.09\Msun$ \citep{gillon16,gillon17,luger17,delrez18,vangrootel18,ducrot18,agol21}. One remarkable feature of TRAPPIST-1 is that the planets are possibly in the longest resonant chain known to date, with period ratios between adjacent planets close to the following ratios (starting with the innermost pair): 8:5, 5:3, 3:2, 3:2, 4:3, and 3:2 \citep{luger17}.

Resonant configurations have been observed in other planetary systems, although they are not the most common orbital configuration \citep{fabrycky14}. 
The formation of mean motion resonances (hereafter MMRs) is often attributed to convergent migration of two planets in a gaseous disc \citep{snellgrove01,lp02}. Furthermore, several exoplanetary systems have been observed in chains of resonances involving three or more planets \citep[see, e.g.,][and the aforementioned TRAPPIST-1 system]{rivera10,gozdziewski16,macdonald16,mills16,christiansen18,leleu21}. Three-body resonances are often referred to as Laplace resonances, in reference to the orbital configuration of the Jovian satellites Io, Europa, and Ganymede. Hydrodynamical simulations have long predicted that resonant chains of low-mass planets could form via type-I migration in gaseous discs \citep{mcneil05,cresswell06,tp07,ogihara09,ida10}.

Using formation models based either on pebble accretion or planetesimal accretion, several studies have shown that resonant chains such as the one seen in TRAPPIST-1 can naturally arise during the early stages of formation and migration in a disc \citep{ormel17,schoonenberg19,coleman19,huang21}. As noted by \citet{coleman19}, the majority of pairs end up in first-order resonances (mainly the 2:1, 3:2 and 4:3 MMRs), with a few in second-order resonance (such as the 5:3 MMR which is relevant for this work), but no pair ends up in third-order resonance (which may be relevant for TRAPPIST-1 since the inner pair has a period ratio of $\sim 1.6$, and therefore lies near the 8:5 MMR). Using a simpler approach where planets are assumed to have already formed, and are initially placed just outside their current positions, and where migration forces are only applied to the outermost planet, \citet{tamayo17} showed that disc migration allows the capture of the TRAPPIST-1 planets into their current configuration. Such a configuration is stable on very long timescales, as opposed to the early $N$-body simulations by \citet{gillon17} which typically showed instability on a timescale of 0.5~Myr. Finally \citet[][see also \citealt{brasser19,huang21}]{papaloizou18} studied the migration and tidal evolution of the system, arguing that the planets may have migrated in two distinct groups, re-assembling later on, thanks to outward migration of the inner planets driven by tidal interactions with the central star.

The TRAPPIST-1 system presents strong TTVs (i.e., deviations of the transit times with respect to the strictly periodic Keplerian case). These variations are useful in order to infer the presence of unseen planets \citep{miralda02,schneider04,agol05,hm05,cochran11,ford12,steffen12a} and to provide a dynamical measurement of planetary masses and eccentricities \citep{nesvorny08,lithwick12,hadden17}. This is especially the case for systems harboring planets in or near MMR for which the variations are strongly enhanced. Based on transit timing data presented in \citet{ducrot2020}, a refined TTV model of the TRAPPIST-1 system was recently published by \citet{agol21}, allowing for an improvement of the mass estimates of the seven planets.

In this paper we use the new analysis from \citet{agol21} to better constrain the formation and dynamics of the TRAPPIST-1 system.  Whether all the planet pairs are indeed in resonances, and how did these resonances form are still open questions. In Sect.~\ref{sec:trappist} we carry out a frequency analysis of the best-fit solution from \citet{agol21} and characterize the resonant nature of the solution. In Sect.~\ref{sec:mig} we explore whether this resonant configuration can be produced by disc migration. Our results are discussed in Sect.~\ref{sec:discussion} and summarized in Sect.~\ref{sec:conclusion}.

\section{Dynamics of TRAPPIST-1}
\label{sec:trappist}

In this section, we study in detail the resonant chain of TRAPPIST-1 based on the latest model inference obtained by \citet{agol21}. The masses and orbital properties of the system are listed in Table~\ref{table:obs}. By means of frequency analysis, we also provide a deep analysis of the periods shown by the TTVs to explore how they are connected to the dynamical evolution of the system.

\begin{table}
\caption{Best-fit solution for TRAPPIST-1 from \citet{agol21}.}
\label{table:obs}
\centering             
\begin{tabular}{l c c c c c}
\hline 
Planet & $m~(m_\oplus)$ & $P$ (d)& $a$ (au) & $e$ & Period ratio \\ 
\hline
b  &  1.375  &  1.511  &  0.012  &  0.007  &  1.603 (8:5) \\
c  &  1.317  &  2.423  &  0.016  &  0.002  &  1.672 (5:3) \\
d  &  0.390  &  4.050  &  0.022  &  0.007  &  1.507 (3:2) \\
e  &  0.704  &  6.103  &  0.029  &  0.005  &  1.509 (3:2) \\ 
f  &  1.044  &  9.210  &  0.039  &  0.010  &  1.342 (4:3) \\
g  &  1.321  &  12.356  &  0.047  &  0.003  & 1.520 (3:2) \\
h  &  0.327  &  18.779  &  0.062  &  0.004  &  \\
\hline
\end{tabular}
\tablefoot{The columns are the mass $m$, period $P$, semi-major axis $a$, eccentricity $e$, and period ratio between adjacent planets. The period ratio is between the given planet and the one immediately exterior to it.  The mass assumes a stellar mass of 0.0898 $M_\odot$ \citep{Mann2019}. An electronic version of this table containing the full set of orbital elements of each planet is available online at the CDS.}
\end{table}

\begin{figure*}
    \begin{center}
    \includegraphics[scale=0.3]{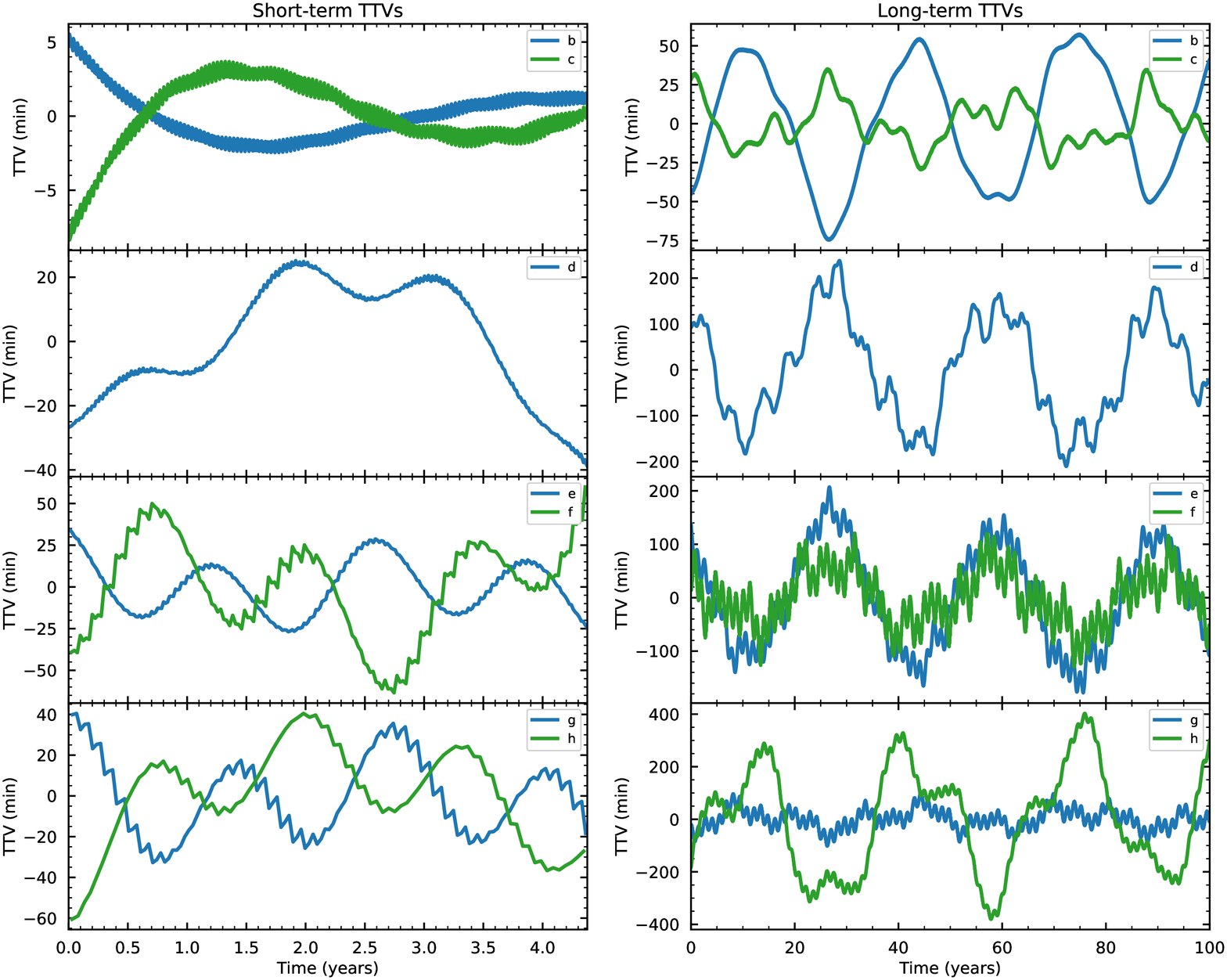}
    \caption{TTVs of the best-fit solution for TRAPPIST-1 from \citet{agol21}. The left panel shows the TTVs over the 1600 days of observation as in Fig.~2 of \citet{agol21}, while the right panel shows the TTVs over 100 yr.}
    \label{fig:agol_ttv}
    \end{center}
\end{figure*}

\begin{figure}
    \begin{center}
    \includegraphics[width=\columnwidth]{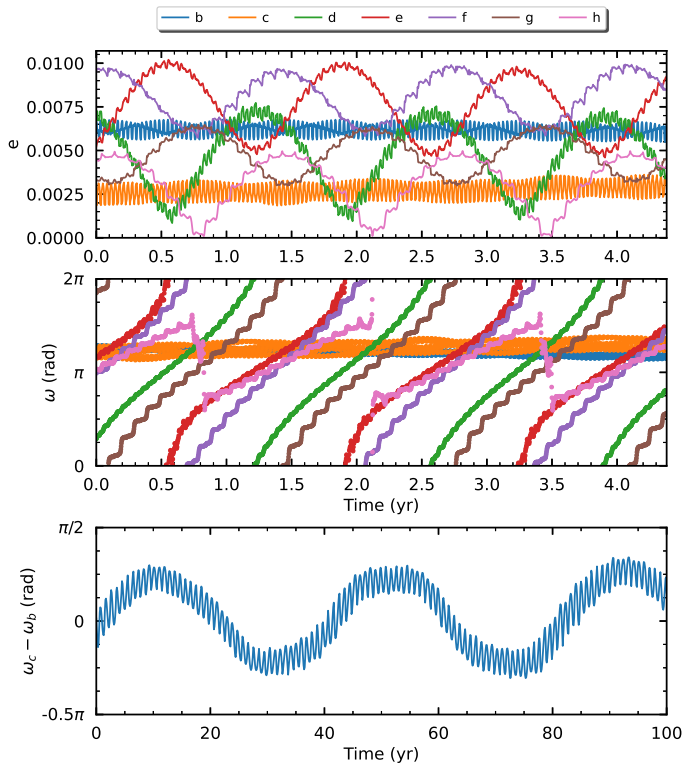}
    \caption{Orbital evolution of the \citet{agol21} best fit: eccentricity in the top panel, argument of pericenter in the middle panel, and difference of arguments of pericenters of planets b and c in the bottom panel (note the different timescale on the bottom panel) .}
    \label{fig:agol_orbit}
    \end{center}
\end{figure}

\begin{figure*}
    \begin{center}
    \includegraphics[scale=0.3]{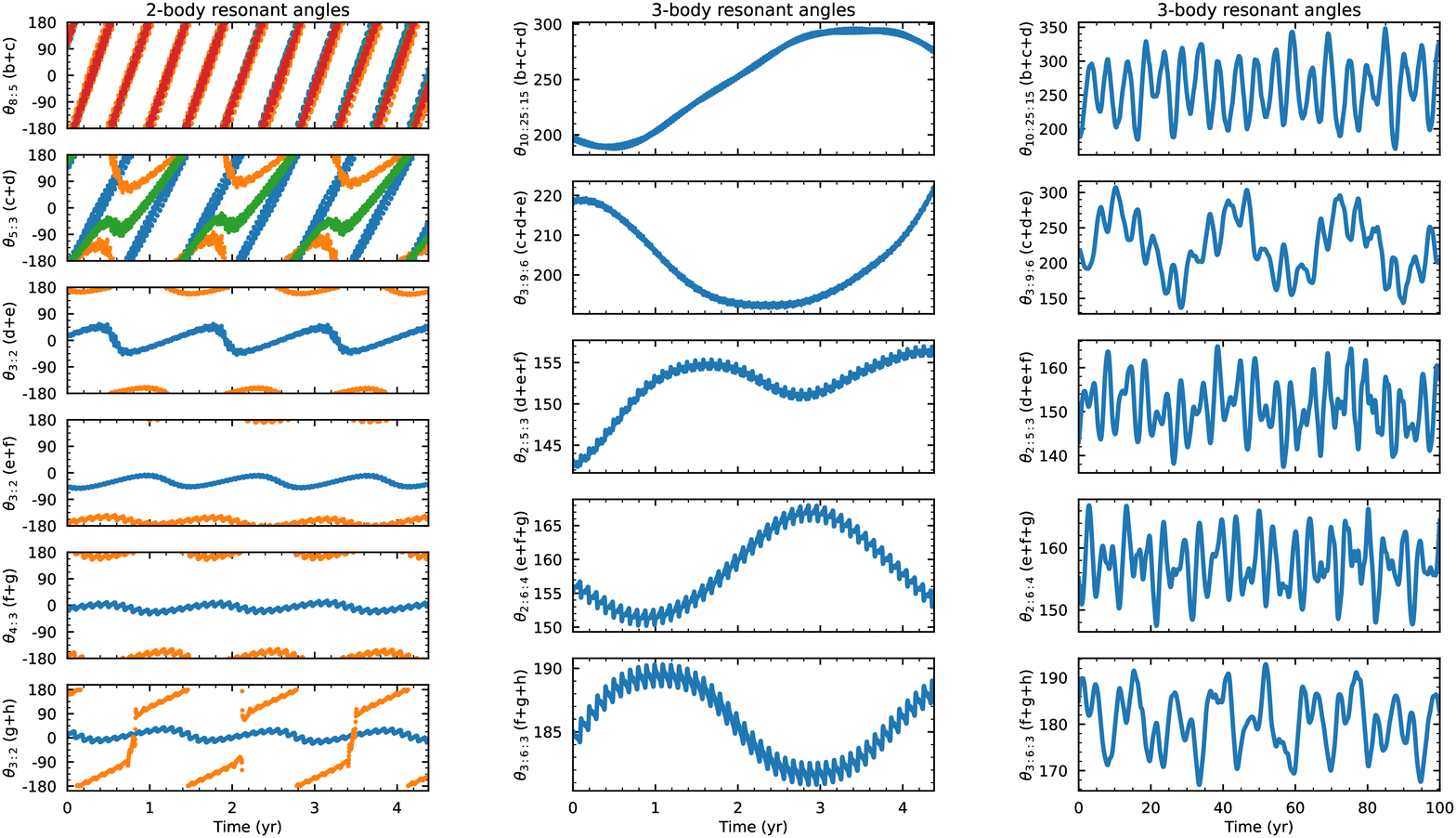}
\caption{Time evolution of the resonant angles of the \citet{agol21} best fit. The left column shows the two-body resonant angles from Eq.~\eqref{eq:res_2b_first} to \eqref{eq:res_2b_last}. The color scheme is as follows: blue is first angle $\theta^{(1)}$, orange $\theta^{(2)}$, and if present, green and red $\theta^{(3)}$ and $\theta^{(4)}$, respectively. The middle and right columns show the three-body resonant angles from \eqref{eq:res_3b_first} to \eqref{eq:res_3b_last}. In the middle column they are plotted over the same 1600~d timescale as the two-body angles. On the right column they are plotted on a longer timeframe of 100~yr.}
    \label{fig:agol_res}
    \end{center}
\end{figure*}

\subsection{Resonant chain}

The seven planets of the recent best-fit solution of \citet{agol21} are all influenced by 2- and/or three-body resonances. The last column of Table~\ref{table:obs} gives the orbital period ratios between adjacent planets, as well as the closest low-order two-body MMRs. The recent best-fit TTV model of \citet{agol21} (see their Fig.~2), based on 4 yr of observation, is reproduced in the left panel of Fig.~\ref{fig:agol_ttv}. Since the planets revolve about the ultra-cool red dwarf star with very short orbital periods (from 1.5 to 19 days), a survey of TRAPPIST-1 on 4 yr was sufficient to clearly see the imprint of the different two-body MMRs between adjacent planets in the TTVs. Indeed, the characteristic timescale of the TTVs (i.e., the period of the sinusoidal TTV signal) for two planets near a $j+k:j$ MMR is \citep[see, e.g,][for first order resonances]{lithwick12}:
\begin{align}
    P_{\rm TTV}=\frac{1}{|j/P_1-(j+k)/P_2|},
\end{align}
with $P_1$ and $P_2$ the orbital periods of the inner and outer planets, respectively. This gives a characteristic period of $\sim 1.3$ yr for the first-order resonances amongst the five outer planets of TRAPPIST-1, which is clearly visible in the TTVs of planets d to h in the left panel of Fig.~\ref{fig:agol_ttv}.
{
For pair b-c, which is near a third-order resonance, the formula gives a period of 0.45~yr. For pair c-d, which is near a second-order resonance, it gives a period of 0.72~yr.
}
However, the TTVs also present several additional shorter and longer periods which we aim to identify precisely in the following based on the analysis of the dynamical evolution of TRAPPIST-1.

Given the data of Table~\ref{table:obs}, the relevant two-body resonant angles are:
\begin{align}
\theta_{4:3}^{(1)} &= 4\lambda_2 - 3\lambda_1 - \varpi_1,\label{eq:res_2b_first}\\
\theta_{4:3}^{(2)} &= 4\lambda_2 - 3\lambda_1 - \varpi_2,\\
\theta_{3:2}^{(1)} &= 3\lambda_2 - 2\lambda_1 - \varpi_1,\\
\theta_{3:2}^{(2)} &= 3\lambda_2 - 2\lambda_1 - \varpi_2,\\
\theta_{8:5}^{(1)} &= 8\lambda_2 - 5\lambda_1 - 3\varpi_1,\\
\theta_{8:5}^{(2)} &= 8\lambda_2 - 5\lambda_1 - 3\varpi_2,\\
\theta_{8:5}^{(3)} &= 8\lambda_2 - 5\lambda_1 - 2\varpi_1 - \varpi_2,\\
\theta_{8:5}^{(4)} &= 8\lambda_2 - 5\lambda_1 - \varpi_1 - 2\varpi_2,\\
\theta_{5:3}^{(1)} &= 5\lambda_2 - 3\lambda_1 - 2\varpi_1,\\
\theta_{5:3}^{(2)} &= 5\lambda_2 - 3\lambda_1 - 2\varpi_2, \\
\theta_{5:3}^{(3)} &= 5\lambda_2 - 3\lambda_1 - \varpi_1-\varpi_2,
 \label{eq:res_2b_last}
\end{align}
for two adjacent planets labelled 1 and 2, with orbital periods $P_1<P_2$, and with $\varpi$ the longitude of pericenter and $\lambda$ the mean longitude. These angles can be combined to form the following three-body resonant angles for the successive triplets of planets:
\begin{align}
&\theta_{10:25:15} = 10\lambda_1-25\lambda_2+15\lambda_3, \label{eq:res_3b_first}\\
&\theta_{3:9:6} = 3\lambda_1-9\lambda_2+6\lambda_3,\\
&\theta_{2:5:3} = 2\lambda_1-5\lambda_2+3\lambda_3,\\
&\theta_{2:6:4} = 2\lambda_1-6\lambda_2+4\lambda_3,\\
&\theta_{3:6:3} = 3\lambda_1-6\lambda_2+3\lambda_3 \label{eq:res_3b_last}.
\end{align}

In order to study further the resonant nature of TRAPPIST-1, we integrate it forward in time, starting from the initial conditions given by the best fit model of \citet{agol21}. Throughout this paper, simulations made use of the \texttt{REBOUND} $N$-body code \citep{rl12}. The simulations were integrated using \texttt{WHFast}, a symplectic Wisdom-Holman integrator \citep{rt15,wh91}. Our integrator includes relativistic corrections as implemented in \texttt{REBOUNDx} \citep{tamayo19}, but neglects any tidal interactions. In Fig.~\ref{fig:agol_orbit}, we show the orbital evolution (the eccentricities and the pericenter arguments) of the 7 planets over the same 1600 days baseline as \citet{agol21}. The eccentricities clearly exhibit some periodic variations, with a distinctive pattern for planets b and c on one hand, and all the other planets on the other hand. This is more apparent when looking at the arguments of pericenters which have a nearly constant average value for planets b and c, and are circulating for the other planets.  There is also a rapid oscillation for planets b and c which occurs on a timescale of $\sim 11.76$ days which is driven by the 3:2 resonant term; this has a small amplitude given that their period ratio is far from 3/2.

In the left panel of Fig. \ref{fig:agol_res}, we computed the two-body resonant angles based on the proximity to the given MMR shown in Table \ref{table:obs}. The best-fit solution shows a resonant behavior, with most two-body angles librating, apart from the angles associated with the 8:5 MMR between the innermost two planets which are circulating. We have also checked the behavior of the 3:2 resonant angles of this innermost pair, and did not observe a resonant behavior. 
However, while not in MMR, the pair b-c presents an alignment of the pericenters, as shown by the libration of the difference of the pericenter arguments around 0, with a period of 40~yr (bottom panel of Fig. \ref{fig:agol_orbit}).

Some similarities in periodicity are apparent between Figs.~\ref{fig:agol_ttv}, \ref{fig:agol_orbit} and \ref{fig:agol_res}. In particular, the two-planet resonant angles between all pairs except the innermost one show variations with the characteristic period of about 1.3~yr previously discussed, the same period that is observed in the TTVs of planets d to h.

The three-body resonant angles quoted in Eqs.~\eqref{eq:res_3b_first}--\eqref{eq:res_3b_last} are also shown in Fig. \ref{fig:agol_res}, on the same timescale of 1600 days as in \cite{agol21} in the middle panel and on a longer timescale of 100~yr in the right panel. As previously noted by \cite{libert13}, the dynamics of three-planet (or Laplace) resonances occurs on a timescale longer than that of two-planet MMRs. The resonant behavior shown in Fig.~\ref{fig:agol_res} suggests that the system is dominated by two-body resonances over the observed baseline of 1600 days, while it is dominated by three-body resonances over longer timescales of over a decade. Moreover, we observe that all the three-body resonant angles of the best-fit solution librate. As expected, they appear nearly constant on the 4 yr of observation, while they present interesting periods on their long-term evolution. 

In light of the above, we computed the TTVs of the best-fit solution of TRAPPIST-1 on a timescale of 100~yr in the right-hand side of Fig.~\ref{fig:agol_ttv}. We will further explore the periodicity of the ``short-term'' and ``long-term'' TTV signals in the next section.

\subsection{Frequency analysis}
\label{sec:freq}

In order to further investigate the dynamics of the system, we carried out a frequency analysis of the best-fit solution. The frequency analysis of the TTVs and resonant angles was done using the method outlined in \citet{laskar93}. This method approximates a periodic signal as a finite sum of trigonometric functions, and allows for a very accurate computation of the amplitude, frequency, and phase of each component of the approximate signal. 

Since the three-body resonances impact the TTVs on timescales longer than two-body resonances \citep{libert13}, we conducted the frequency analysis on two datasets: the ``short-term'' dataset of 1600 days, and the ``long-term'' dataset of 100~yr. By carrying out a frequency analysis on the two datasets, we can extract the relevant periods of the TTV signal of each planet. In total, nine periods were unveiled, most of them being present in the TTV signal of each planet  but with different amplitudes. They are listed in the first column of Table~\ref{table:ttv_all}.

\begin{table*}
    \caption{Main periods (in years) found by frequency analysis in the TTVs and the resonant angles.}
\label{table:ttv_all}
    \centering
\begin{tabular}{ccccccc|ccccc}
\hline
 TTV periods (yr)& $\theta_{\rm bc}$ & $\theta_{\rm cd}$ & $\theta_{\rm de}$ & $\theta_{\rm ef}$ & $\theta_{\rm fg}$ & $\theta_{\rm gh}$ & $\theta_{\rm bcd}$ & $\theta_{\rm cde}$ & $\theta_{\rm def}$ & $\theta_{\rm efg}$ & $\theta_{\rm fgh}$ \\
\hline
0.032 & \checkmark & \checkmark & & & & & & & & & \\
0.09   & & & \checkmark & \checkmark & \checkmark & \checkmark & & & & & \\
0.45 & \bcheckmark &  & & & & & & & & & \\
0.68 & & \bcheckmark & & & & & & & & & \\
1.3   & & & \bcheckmark & \bcheckmark & \bcheckmark & \bcheckmark & & & & & \\
\hline
3.3   & & & & & & & \checkmark  & \checkmark  & \bcheckmark  & \bcheckmark  & \checkmark \\
5.1   & & & & & & &   \bcheckmark  & \checkmark  & \checkmark & \checkmark & \checkmark\\
12.3  & & & & & & & \checkmark & \checkmark  & \checkmark & \checkmark  & \bcheckmark\\
31.5 &  & & & & & & \checkmark  & \bcheckmark & \checkmark  & \checkmark  & \checkmark\\
\hline
\end{tabular}
\tablefoot{ Ticks indicate that the period is present in the signal with a large amplitude. Circled bold ticks highlight the periodic component with the largest amplitude.}
\end{table*}

Regarding the short-term dynamics of the TTVs, the signals of the outer five planets are dominated by a periodic modulation of 1.3~yr, as can be seen on the left-hand side column of Fig. \ref{fig:agol_ttv}. An additional very short period of 0.09~yr is also present, and particularly visible in planets e to h. The importance of these two periods will be further discussed in Section~\ref{sec:discussion}. Planets b and c, on the other hand, are mostly dominated by a short period variation which is three times faster, at 0.032~yr (corresponding to the $\sim 11.76$~d period seen in the eccentricities and arguments of pericenter of planets b and c), as well as an additional period variation of 0.45~yr.  

The long-term TTVs mainly vary on a period of $31.5$~yr for all the planets, as visible on the right-hand side column of Fig. \ref{fig:agol_ttv}. Planets d to g possess a common period of 3.3~yr. A common period of 5.1~yr is shared across all the planets. As also visible in the last panel of Fig.~\ref{fig:agol_ttv}, a period of 12.3~yr is also reported by the frequency analysis.

Table \ref{table:ttv_all} shows the main periods found in the time series of the resonant angles, for the two-body resonances between adjacent pairs in columns 2 to 7 and for the three-body resonances in columns 8 to 12. A tick indicates that the period is present in the signal with a large amplitude. A circled  bold tick indicates the periodic component with the largest amplitude. This table allows one to immediately visualize two groups: two-body angles are associated with short periods and three-body angles to long periods. In particular, the period of 1.3~yr is clearly visible in the eccentricity, TTVs and two-body resonances, pointing towards a chain of two-body resonances as the origin of this signal. On the other hand, periods of 31.5, 12.3, 5.1, and 3.3~yr modulate the signal on long timescales, both in TTVs and three-body resonant angles.

Finally, we note that we explored 100 draws from the posterior distribution of \citet{agol21}. They all exhibit the same resonant structure, and in all of them we recovered the main periods of 1.3 and 31.5~yr. The angles associated with the 8:5 MMR between planets b and c was also circulating in all of the simulations.  

Our main conclusion is that on short timescales, the main frequencies of the TTVs derived from the best-fit solution arise from two-body resonances, while they arise from three-body resonances on longer timescales. Our analysis is relevant for follow-up observational campaigns of TRAPPIST-1. In the coming decade, enough data could be collected to refine the best-fit solution. Our analysis suggest that decade-long observational span could give enough time to see the 3.3 and 5.1~yr frequencies arise. More measurements would also help regarding the peculiar dynamics of the inner pair, since its currently observed period ratio of 1.6 is hard to reproduce through disc migration (see Sect.~\ref{sec:mig}). Hence, in addition to being relevant to future observations, our analysis puts constraints on the physical processes that took place during the formation and migration of TRAPPIST-1. In the next section, we explore whether the resonant chain of TRAPPIST-1 can be formed via smooth disc migration.

\section{Formation of TRAPPIST-1 by disc migration}
\label{sec:mig}

In this section, we study the late-stage formation of TRAPPIST-1, in particular the migration phase. We aim to see whether the resonant chain of TRAPPIST-1 discussed in the previous section can be reproduced by disc migration.   

\subsection{Disc model}
\label{sec:disc}

Given the masses of the TRAPPIST-1 planets, they are unlikely to be in the gap-opening (i.e., Type-II) migration regime. The planet-to-star mass ratios in the TRAPPIST-1 system range from $\sim10^{-5}$ to $\sim5\times10^{-5}$. Around a Sun-like star, this would correspond to a super-Earth to Neptune size planet. In this mass regime, we follow \citet{papaloizou18} and apply semi-major axis and eccentricity damping forces corresponding to Type-I migration. We use the forces prescribed by \citet{cresswell08} which we implemented in \texttt{REBOUNDx} \citep{tamayo19}. We assume that the disc mass remains constant throughout the simulations.

Let us consider a planet of mass $M_{\rm p}$ orbiting a star of mass $M_*$, at a distance $r_{\rm p}$ and with a Keplerian frequency $\Omega_{\rm p}$. The planet is embedded in a disc of surface density $\Sigma$ and aspect ratio $H/r$. We define $\beta=-\id\ln\Sigma/\id\ln r$ the slope of the surface density. Following \citet{tanaka02}, we define the wave timescale as:

\begin{align}
    \tau_{\rm w} = \frac{M_*}{M_{\rm p}}\frac{M_*}{\Sigma(a_{\rm p})a_{\rm p}^2}\left( \frac{H}{r} \right)^4 \Omega_{\rm p}^{-1}.
\end{align}
Hydrodynamical simulations by \citet{cresswell08} indicate that the semi-major axis then decays at a rate given by
\begin{align}
\label{eq:ta}
    \tau_a = \frac{\tau_{\rm w}}{2.7+1.1\beta}\left( \frac{H}{r} \right)^{-2}P(e_{\rm p})
\end{align}
with 
\begin{align}
    P(e) = \frac{1+\left(\frac{e}{2.25(H/r)}\right)^{1.2}+\left(\frac{e}{2.84(H/r)}\right)^{6}}{1-\left(\frac{e}{2.02(H/r)}\right)^{4}}.
\end{align}
Note that if the surface density increases sharply with radius (as would for instance be the case at the inner edge of the disc), one would have $\beta<0$. This raises the possibility that the term $2.7+1.1\beta$ is negative, hence halting or even reverting the direction of migration of the planet as it reaches the inner edge \citep[see, e.g.,][]{kretke12}. Similarly, this formula indicates that the migration can also be outward for $e>2.02H/r$ \citep{cresswell08}. Finally, the eccentricity is damped at a rate: 
\begin{align}
    \label{eq:te}
    \tau_e = \frac{\tau_{\rm w}}{0.78} F(e_{\rm p})
\end{align}
with
\begin{align}
    F(e) = 1-0.14\left(\frac{e}{(H/r)}\right)^2+0.06\left(\frac{e}{(H/r)}\right)^3.
\end{align}

\subsection{Halting Type-I migration at the inner edge}
\label{sec:edge}

\begin{figure}
    \begin{center}
    \includegraphics[width=\columnwidth]{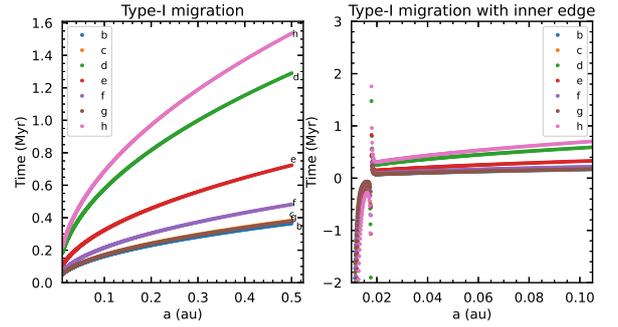}
    \caption{Migration time of the TRAPPIST-1 planets in a simple power-law disc (left), and in a disc presenting an inner edge (right) such as the one prescribed by Eq. \eqref{eq:sigmatap}. The presence of an inner edge slows down migration, and can even reverse it (corresponding to negative migration times).}
    \label{fig:typeI}
    \end{center}
\end{figure}

\begin{figure}
    \begin{center}
    \includegraphics[scale=0.3]{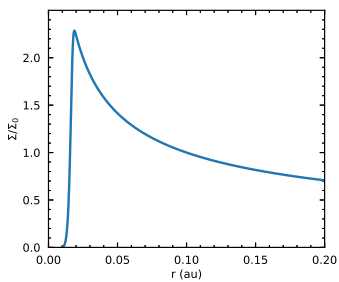}
    \caption{Surface density profile given by Eq. \eqref{eq:sigmatap} used in our simulations in order to stop migration at the inner edge.}
    \label{fig:surface_density}
    \end{center}
\end{figure}

\begin{figure*}
    \begin{center}
    \includegraphics[width=2\columnwidth]{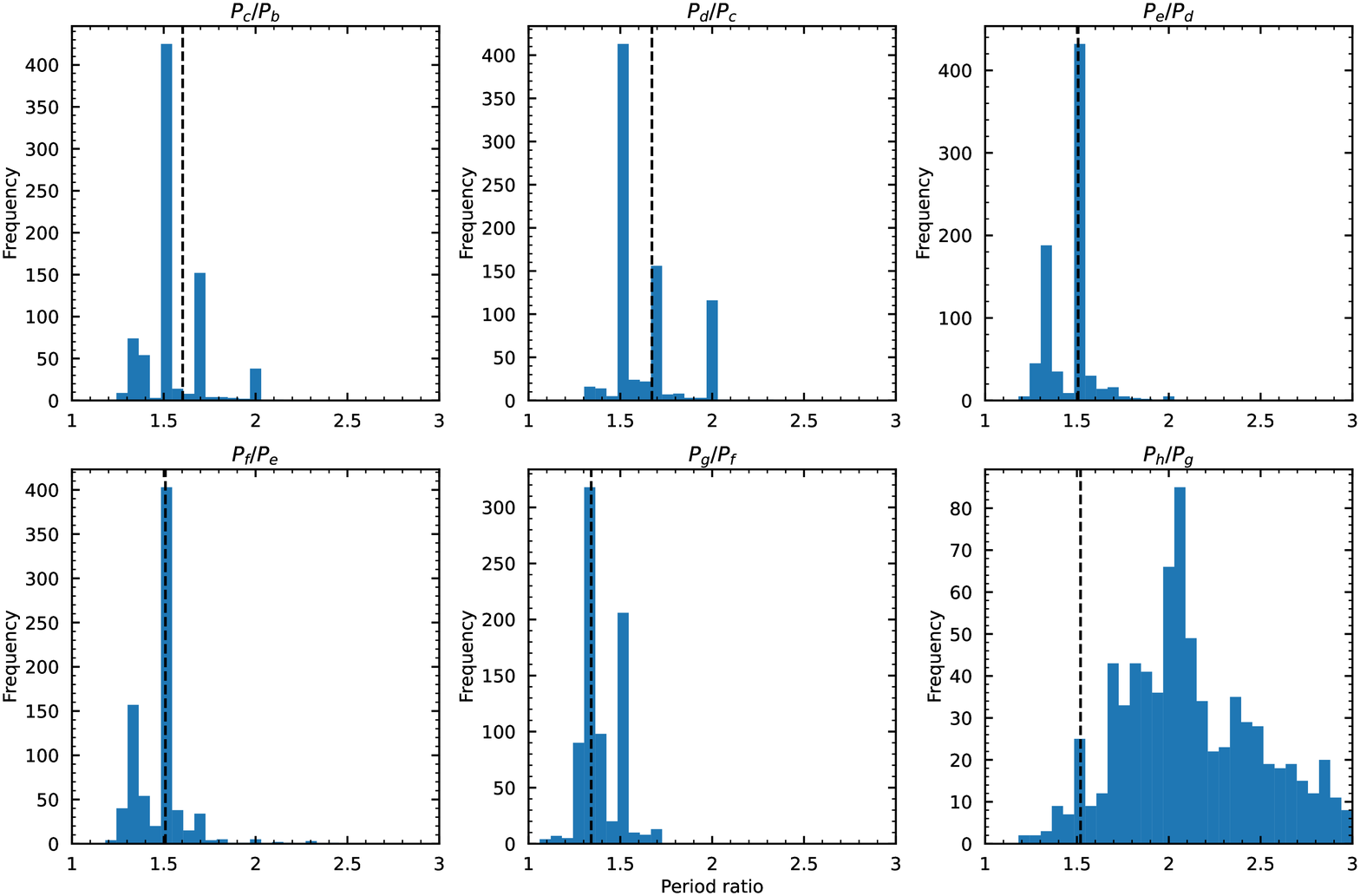}
    \caption{Final period ratio found in the 792 simulations of disc migration for the TRAPPIST-1 planets, achieved using the initial conditions described in Sect.~\ref{sec:ic}. The dashed vertical line in each panel represents the currently observed period ratio of the given pair (see Table \ref{table:obs}).}
    \label{fig:m3}
    \end{center}
\end{figure*}

The Type-I migration timescale defined in Eq.~\eqref{eq:ta} is inversely proportional to the planet mass. All other things being equal, more massive planets migrate faster. Resonant capture of two migrating planets requires a convergent migration, that is, the outer planet needs to migrate faster than the inner one. Planets b and c have very similar masses (see Table \ref{table:obs}), and their relative migration speed is slow. Although planet b is slightly more massive, intricate details of the formation process and disc structure allow for the possibility of convergent migration. More interestingly, planet d is significantly less massive than planets b and c. Assuming a smooth disc described by a power-law all the way to the inner edge, Eq. \eqref{eq:ta} indicates that planet d never migrates fast enough to catch up with planets b and c and capture planet c in the 5:3 MMR. This can be seen on the left-hand side panel of Fig.~\ref{fig:typeI}, where we plot the migration time of all seven planets, assuming they migrate in a disc with surface density $\Sigma=\Sigma_0 r^{-1/2}$, where $\Sigma_0$ is calculated assuming the disc mass is $\Md=10^{-4}~\Msun$ between an inner edge $\rin=0.01$~au and outer edge $\rout=5$~au. It is clear that planet d migrates much slower than planets b and c. However, planet e, f, and g all migrate faster than the planet immediately interior to it. This raises the possibility that planets d, e, f and g were all captured together in a resonant chain, which would explain why their TTVs share so many common periods, as shown in Sect.~\ref{sec:freq}. Finally, planet h being of much lower mass, it would have arrived later. 

A possible scenario is therefore that planets b and c migrate together until the inner edge of the disc where they stopped migrating. Meanwhile (or perhaps even at a later stage) further out in the disc, planets d to g migrated, with migration speed increasing for further out planets, causing them to form one long chain (or two small chains). They finally encountered planets b and c whose migration was already halted at the inner edge, captured them in a resonance, and settled at the inner edge of the disc. Eventually planet h came in and captures planet g in a resonance. Note that \citet{papaloizou18} also examined the possibility that TRAPPIST-1 migrated as two subsystems. In their case, it was an outward migration caused by tidal interactions with the central star which allowed the two sub-systems to 're-attach' \citep[see also][for a more complete scenario invoking both disc migration and stellar tides]{huang21}.

In order for the scenario described above to work, one needs a mechanism to halt migration at the inner edge. As previously noted, the term $2.7+1.1\beta$ in the migration timescale given by Eq. \eqref{eq:ta} allows for the migration to be slowed down or even reversed for $\beta<0$, i.e. for positive surface density slopes. A sharp edge at the inner end of the disc can provide such a positive slop. We therefore model the disc surface density profile as a combination of a power-law and an inner edge:
\begin{align}
    \label{eq:sigmatap}
    \Sigma(r) = \Sigma_0 \left(\frac{r}{\rin} \right)^{-s} \tanh\left(\frac{r-0.7\rin}{\rin} \right)^6.
\end{align}
This surface density profile is plotted in Fig.~\ref{fig:surface_density} for $s=1/2$. The corresponding migration timescales (zoomed in on the inner 0.1 au) are shown on the right-hand side of Fig.~\ref{fig:typeI}. As the planets approach the inner edge, their migration speed becomes slower, and can even be reversed (as indicated by the negative migration times). This provides a way to halt migration at the inner edge \citep[see, e.g.,][]{masset06b,ogihara18}.

\subsection{Initial conditions}
\label{sec:ic}
We assume that the planets have fully grown to their presently observed mass at the start of the simulation, although the planets might have continued experiencing mass growth during the later stages of migration \citep[see, e.g,][]{coleman19}. 
We assume the planetary system to be coplanar, and we initiate the planets with very low eccentricities ($10^{-4}$). The initial orbital periods of the planets are selected using the following relation
\begin{equation}
    P_{\rm k, 0} = a^k b_k P_{\rm k, obs}
\end{equation}
where $P_{\rm k, 0}$ is the initial period of planet $k$, $P_{\rm k, obs}$ its currently observed period (see Table~\ref{table:obs}), $a$ is a random number which determines how wide the planets are initiated away from their current period ratio (\citet{tamayo17} used a=1.02), and $b_k$ is a random number between 1.2 and 1.9. Overall, this law causes the planets to start with periods between 1.2 and 2.5 larger than their observed ones. The longitudes of pericenter, longitudes of ascending nodes, and mean anomalies are all randomly selected between 0 and $2\pi$.
Relativistic precession is included, as implemented in \texttt{REBOUNDx}.

Regarding our disc model, we chose the disc mass randomly from a uniform distribution between $\Md=10^{-5}~\Msun$ and $\Md=10^{-4}~\Msun$, with a radial extension between an inner edge $\rin=0.01$ and an outer edge $\rout=5$ au, following the estimate provided by \citet{papaloizou18}. We also assume that the index of the power-law part of the surface density in Eq.~\eqref{eq:sigmatap} is either $s=1/2$ or $s=1$. We pick $H/r$ from a uniform distribution between 0.025 and 0.035. As noted by \citet{cresswell06}, it is sometimes necessary to multiply the eccentricity damping timescale given by Eq. \eqref{eq:te} by a numerical factor $Q_{\rm e}$ to provide a better match between the linear theory and the simulations. They suggested $Q_{\rm e}=0.1$. In preliminary $N$-body simulations, we found that in order to obtain eccentricities similar to the ones observed in TRAPPIST-1, low values of $Q_{\rm e}$ were necessary, suggesting a strong eccentricity damping. We chose $Q_{\rm e}$ from a uniform distribution between 0.02 and 0.1.

\subsection{Results}
\label{sec:results}
In this section we present the results of the simulations of migrating planets in a disc, aimed at reproducing the architecture of TRAPPIST-1. In Fig.~\ref{fig:m3} we display the outcomes of an ensemble of 1000 simulations carried out using the initial conditions described in Sect.~\ref{sec:ic}. We plot the final period ratios of adjacent pairs of planets, as well as the observed period ratios. Apart from pairs b-c and g-h, all pairs of planets are routinely found in configurations matching their observed ones (dashed lines in Fig.~\ref{fig:m3}). The planets c-d are regularly captured in the 5:3 MMR (although the most frequent capture is in the 3:2 MMR), d-e and e-f both in the 3:2 MMR, and f-g in the 4:3 MMR (although many pairs also end in the 3:2 MMR), as found in the observations.
The slow migration of planet h, as detailed in Sect.~\ref{sec:edge}, combined with our finite integration time, leads to many simulation outcomes where planet h is not captured in any resonance, although we observe pile-ups at the 2:1 and 3:2 MMRs with planet g.

Regarding the inner planets b and c, most pairs are trapped in the 3:2 or the 5:3 MMR, with additional pairs captured in the 2:1 MMR. Seven simulations showed pairs b-c having a period ratio of 1.6. Interestingly, in these seven simulations, the pair c-d was also captured in the observed 5:3 MMR, suggesting that three-body resonances play a significant role during the migration \citep[see also][]{charalambous18}. In addition, all the simulations that led to planets b and c having a period ratio of 1.6 had a similar disc mass of $\Md\approx 3.5\times10^{-5}~\Msun$.

\subsection{Analysis of one simulation}
\label{sec:best_disc}

\begin{figure}
    \begin{center}
    \includegraphics[scale=0.3]{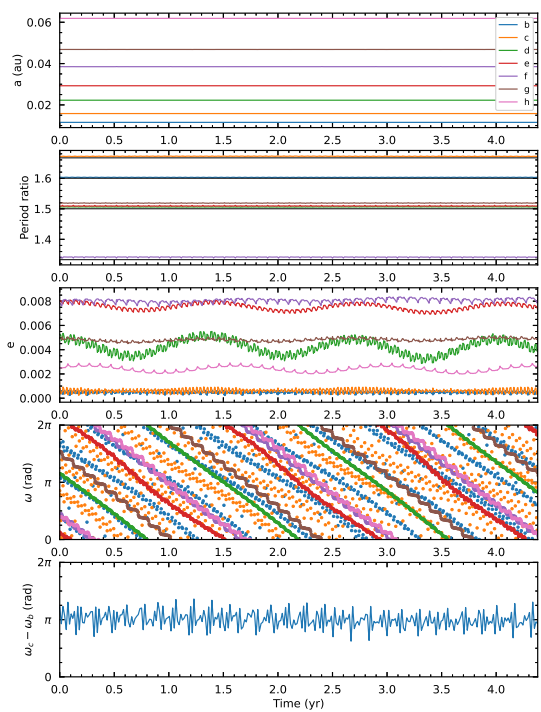}
    \caption{Final orbital configuration of one of the simulations described in Sect.~\ref{sec:ic}. From top to bottom: semi-major axes, period ratios, eccentricities, arguments of pericenters, and difference of arguments of pericenters between planets b and c.} 
    \label{fig:m6r72_orb}
    \end{center}
\end{figure}

\begin{figure*}
    \begin{center}
    \includegraphics[scale=0.3]{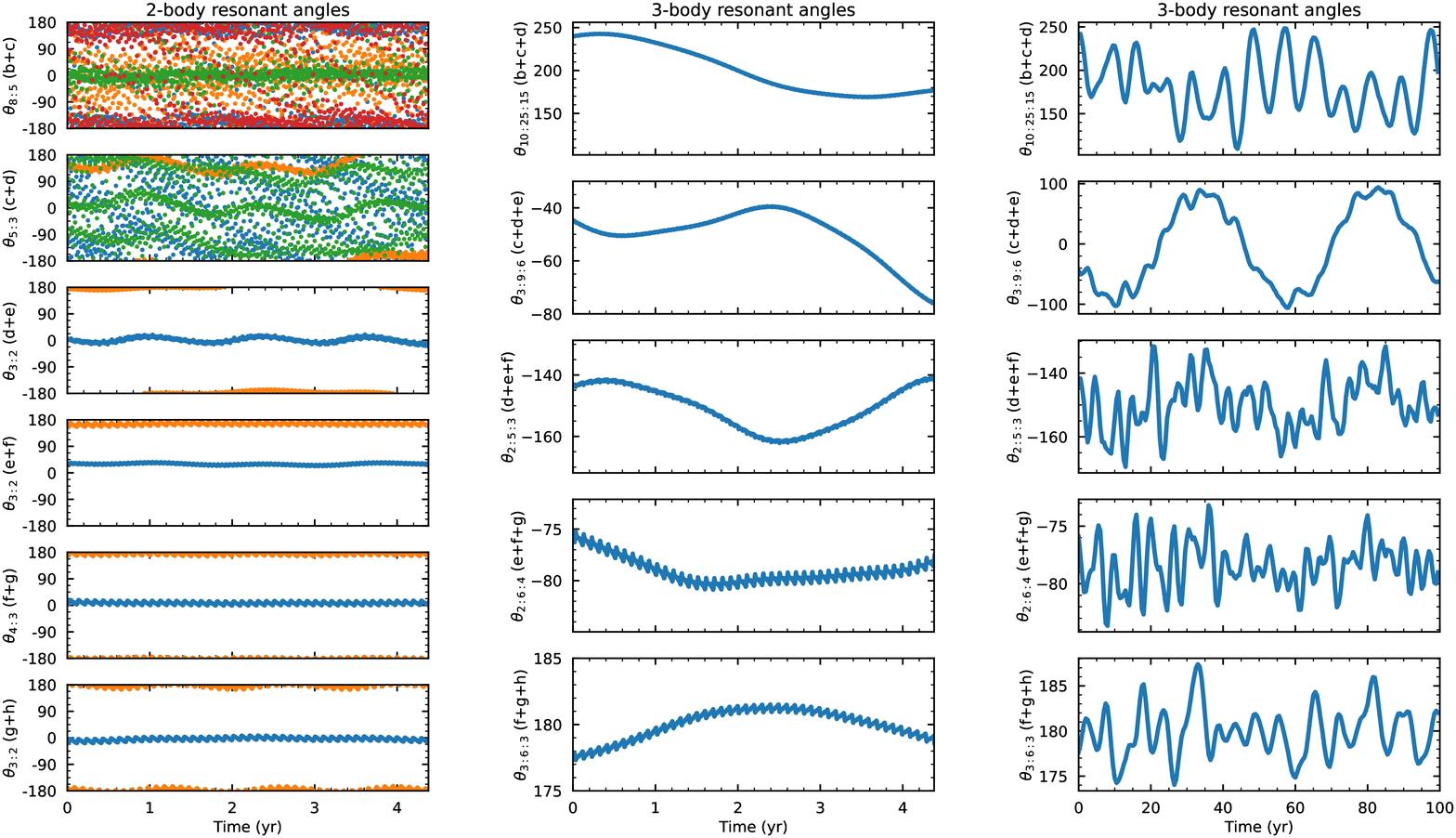}
    \caption{Resonant state of the simulation shown in Fig. \ref{fig:m6r72_orb}. The columns are the same as in Fig. \ref{fig:agol_res}.} 
    \label{fig:m6r72_res}
    \end{center}
\end{figure*}

We now focus on one particular simulation where the period ratio between planets b and c was locked around 1.6. 
For this simulation, the disc parameters are $H/r=0.0344$, $s=0.5$, $\Sigma_0=2.386\times10^{-6} \Msun\ {\rm au}^{-2}$, and $Q_e=0.0134$.
In this particular simulation, planet h was not captured in the 3:2 MMR with planet g, so we restarted a series of simulations with the same initial conditions but a slightly different initial semi-major axis for planet h. We present in Fig. \ref{fig:m6r72_orb} one such simulation where planet h was captured in the 3:2 MMR with planet g.
As can be seen on Fig. \ref{fig:m6r72_orb}, the final semi-major axes and period ratios are very close to the observed ones. The eccentricities are also in the range of observed values, albeit with smaller amplitude compared to the best-fit solution presented in Fig. \ref{fig:agol_orbit}. We also point out that the angular momentum deficit \citep[AMD, see][]{laskar97} of this simulated system differs only by 5\% compared to the AMD of the observed TRAPPIST-1 system\footnote{The AMD of the observed system was computed assuming that the system is coplanar.}.
One major difference with the best-fit solution shown in Fig.~\ref{fig:agol_orbit} arises when considering the arguments of pericenter. In the best-fit case, planets b and c precess at much lower rates than the rest of the system, and their difference in arguments of pericenters oscillates around 0 with a period of about 40~yr. Regarding the system arising from disc migration, the arguments of pericenter are now all circulating at the same rate, but the difference of arguments of pericenters of the pair b-c is librating around $\pi$ .

We now examine the resonant state of the simulated system, shown in Fig.~\ref{fig:m6r72_res}. Regarding planets d to h, all the two-body and three-body resonant angles show very similar trends, compared to the best-fit solution shown in Fig. \ref{fig:agol_res}. Although the amplitudes of libration slightly differ, the main frequencies appearing in the two-body MMRs in Fig. \ref{fig:agol_orbit} are also visible in Fig. \ref{fig:m6r72_res}. The evolutions of the three-body angles are similar on a short timescale (note that the center of libration of the angles associated with triplets d-e-f and e-f-g are of the reversed sign), while on longer timescales they reveal a main period of 40~yr, slightly longer than the 31.5~yr period of the best-fit solution.

\begin{figure}
    \begin{center}
    \includegraphics[scale=0.3]{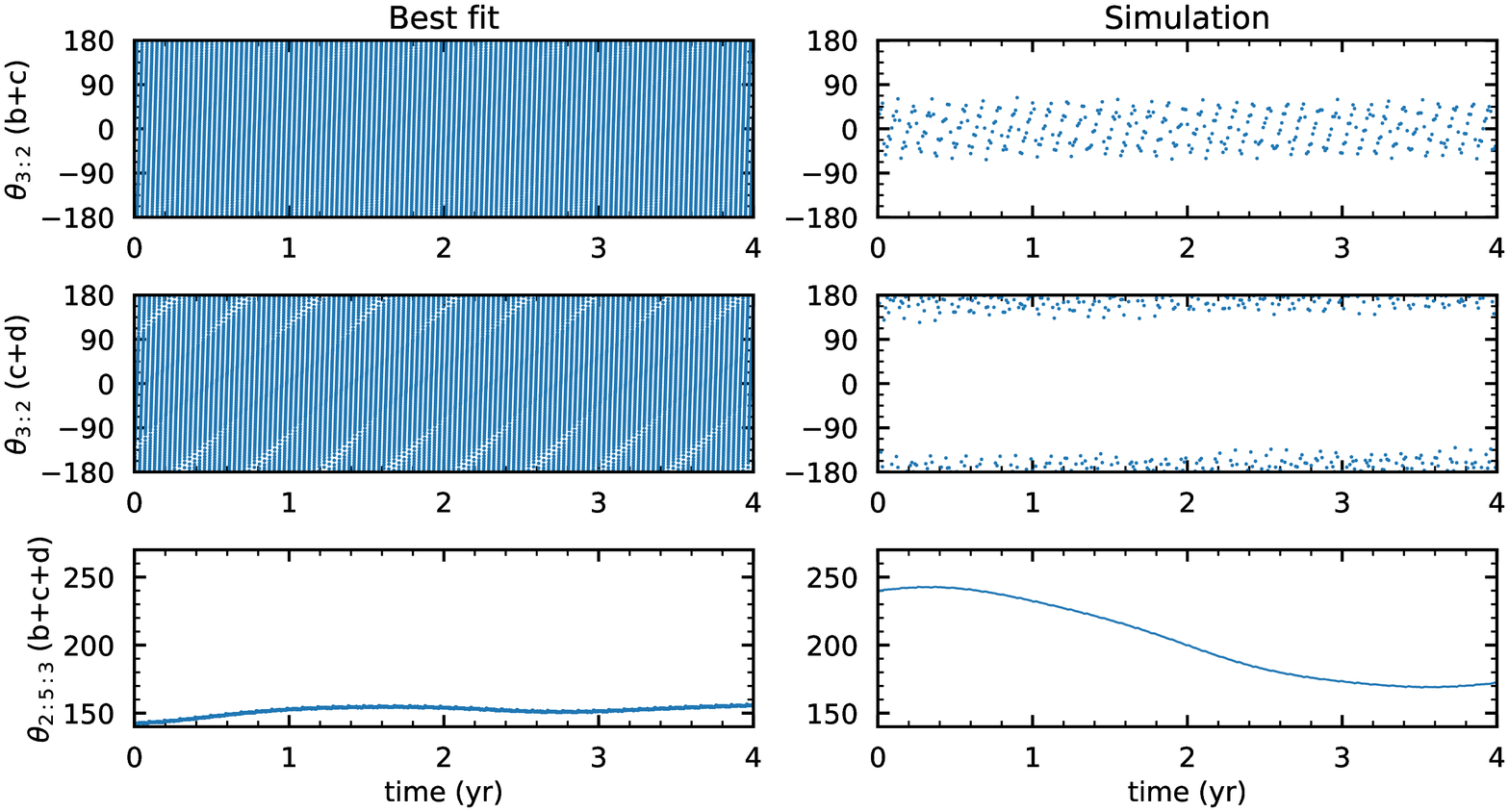}
    \caption{Time evolution of the resonant angles related to the 3:2 MMR between planets b and c (top panel), c and d (middle panel), and their combination as a Laplace angle (bottom panel), for the best-fit solution (left column) and the illustrative simulation of Fig.~\ref{fig:m6r72_orb} (right column).}
    \label{fig:32}
    \end{center}
\end{figure}

Differences arise when considering the two-body angles associated with the 8:5 MMR between planets b and c, which are in a state of clear circulation in the best-fit case. In the disc-migration simulation, although they spend a large fraction of their time around 0 or $\pi$, they are not strictly librating, but rather circulating.  
Similarly, in the system formed by disc migration, planets c and d do not appear to be as deep in the 5:3 MMR as in the best-fit case. In both cases, the angle $\theta_{5:3}^{(2)} = 5\lambda_{\rm d} - 3\lambda_{\rm c} - 2\varpi_{\rm d}$ is librating. However the two other angles behave differently in both sets of simulations due to the different precession rate of $\varpi_{\rm c}$. In Fig.~\ref{fig:32}, we compare the evolution of the angles associated with the 3:2 MMR for pairs b-c and c-d in both sets of simulations, as well as the associated Laplace angle. While the 3:2 angles are not librating in the best-fit case, they are librating with large amplitude in the disc-migration simulation. This libration is only made possible because the arguments of pericenters of planets b and c are precessing much faster in the disc-migration case.

\begin{figure}
    \begin{center}
    \includegraphics[scale=0.3]{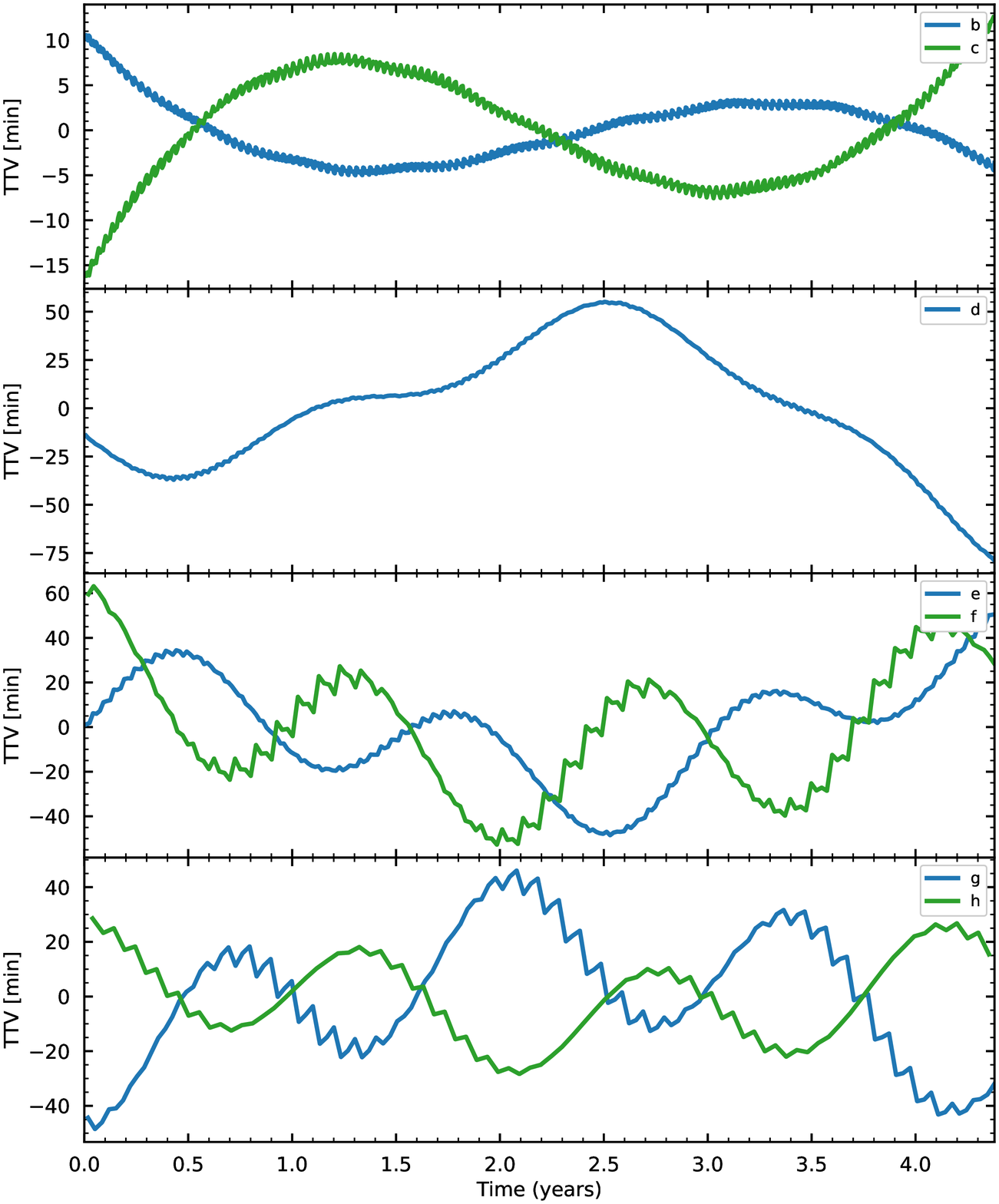}
    \caption{TTVs computed at the end of the simulation shown in Fig. \ref{fig:m6r72_orb}.} 
    \label{fig:m6r72_ttv}
    \end{center}
\end{figure}

Finally, the TTVs associated with this simulation are shown in Fig.~\ref{fig:m6r72_ttv}. The TTVs of the system formed by smooth migration in a disc reproduce very well the ones of the observed TRAPPIST-1 system. Comparing Figs.~\ref{fig:m6r72_ttv} and \ref{fig:agol_ttv}, we recover the very fast variations in the TTVs of planets b and c, as well as the 0.09 and 1.3~yr signals in the outer planets. In addition, the amplitudes of the TTVs are similar. As was previously highlighted in \citet{teyssandier20} for the K2-24 planetary system, we have shown that it is possible to recover most of the orbital architecture and TTV signal of TRAPPIST-1 from smooth disc migration.

\begin{figure}
    \begin{center}
    \includegraphics[scale=0.32]{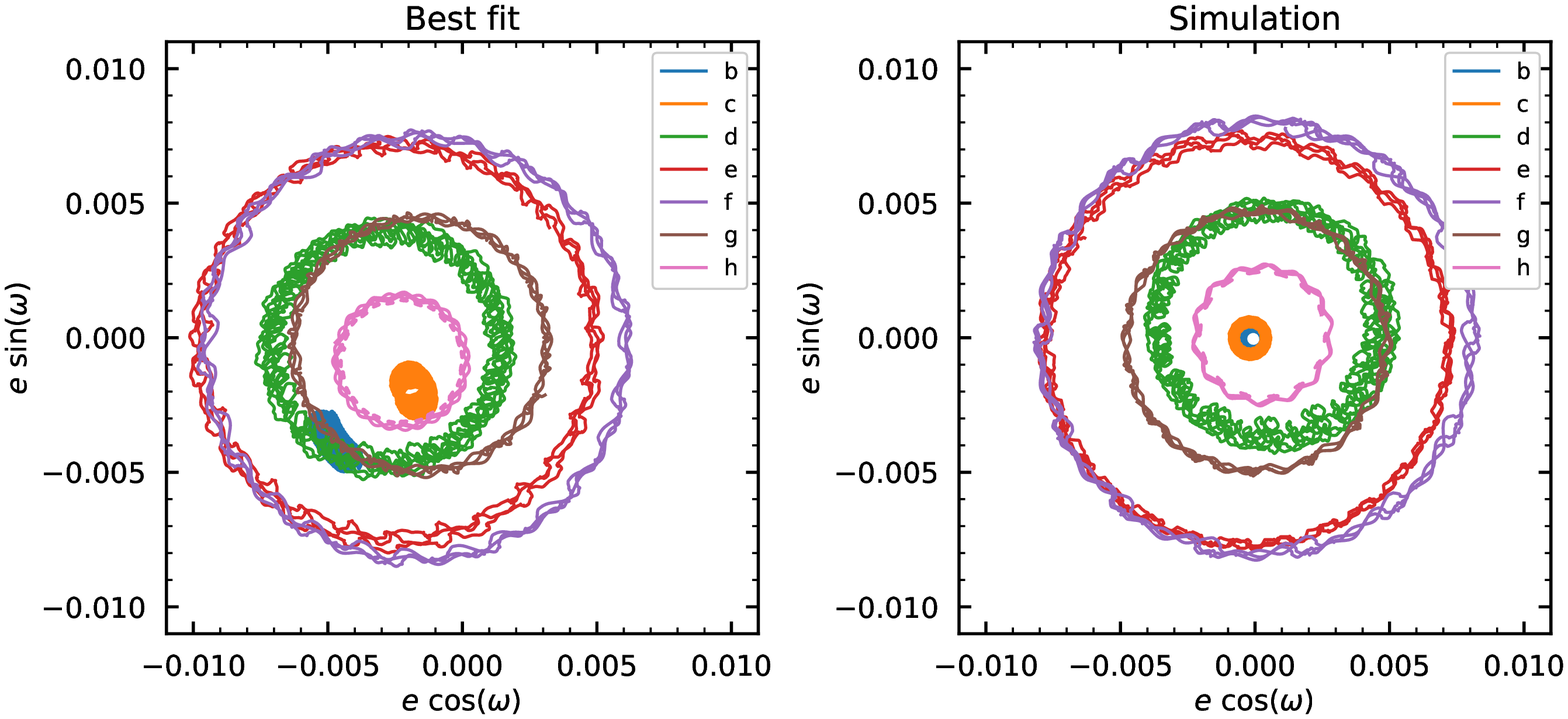}
    \caption{Dynamical evolution of the TRAPPIST-1 system represented in the $e\cos\omega\textrm{--} e\sin\omega$ plane. {\it Left}: evolution of the best-fit solution from \citet{agol21}. {\it Right}: evolution of the system obtained through disc migration. In both panels, the system is represented over the course of 4~yr.
    }
    \label{fig:kh}
    \end{center}
\end{figure}

There is a good agreement between the TTV signals of the best-fit solution and the disc-migration simulation, despite the fact that the resonant dynamics of the pairs b-c and c-d differ between the two cases. In order to understand this discrepancy, we examine the trajectories of the planets in the  $e\cos\omega\textrm{--} e\sin\omega$ plane (see Fig.~\ref{fig:kh}). It is clear from this figure that the architecture of the outer system (planets d to h) is very similar between the best-fit solution and the migration simulation. Although the centers are slightly offset, in both cases the trajectories follow annuli of similar radius and thickness. The forced eccentricities of the outer planets are driven by their captures into first-order resonances. Their free eccentricities have been damped during disc migration and are therefore expected to be small. As was previously noted, the main difference is in the trajectories of planets b and c. In the migration simulation they follow small thick annuli centered on $(0,0)$, meaning that their free eccentricities are very small compared to their forced eccentricities. In the best-fit case, the precession of planets b and c is very slow, and both planets have a larger value of free eccentricity than forced eccentricity. Thus, over the 4~yr window represented in Fig.~\ref{fig:kh}, the trajectories of these two planets are off-centered from the origin. This raises the question of why the TTVs are similar in both cases. As noted by \citet{lithwick12} and \citet{linial18}, TTVs are mostly influenced by the forced eccentricities and a linear combination of free eccentricities, which are similar in both sets of simulations. In the best-fit case, the free eccentricities of planets b and c are offset by a comparable amount, so that the trajectories of their eccentricity vectors maintain similar behaviors on long timescales. Thus, it is difficult to constrain the absolute values of the free eccentricities of each planet, while the relative values are better constrained. Hence the planets still experience the same proximity at conjunctions, and thus feel the same kicks, resulting in the same TTV pattern.

\section{Discussion}
\label{sec:discussion}

The numerical experiments presented in Sect.~\ref{sec:results} show that under generic initial conditions, forming a system like TRAPPIST-1 through disc migration alone is a rare event. Since only a handful of our simulations gave the correct period ratios for pair b-c, it is impossible to estimate the occurrence rate of systems where the angles associated with its 8:5 MMR librate or circulate. We therefore ran 200 additional simulations similar to the ones presented in \citet{tamayo17}: all planets pairs start within 2\% of their observed period ratios, and only the outermost planet migrates with a fixed timescale of $3\times10^4$~yr (eccentricity damping is applied to all planets). For planets b-c, we find that 92\% of the simulations show a libration of the angles associated with the 3:2 MMR (even though their period ratio is $\sim$~1.6). The rest show libration of the 8:5 resonant angles. Among those, some even show a very slow variation of the arguments of pericenter of planets b and c (as is observed in the best-fit solution). Regarding planets c-d, 46\% show libration of the 3:2 MMR angles, while 54\% show libration of the 5:3 MMR angles. 

\begin{figure}
    \begin{center}
    \includegraphics[scale=0.4]{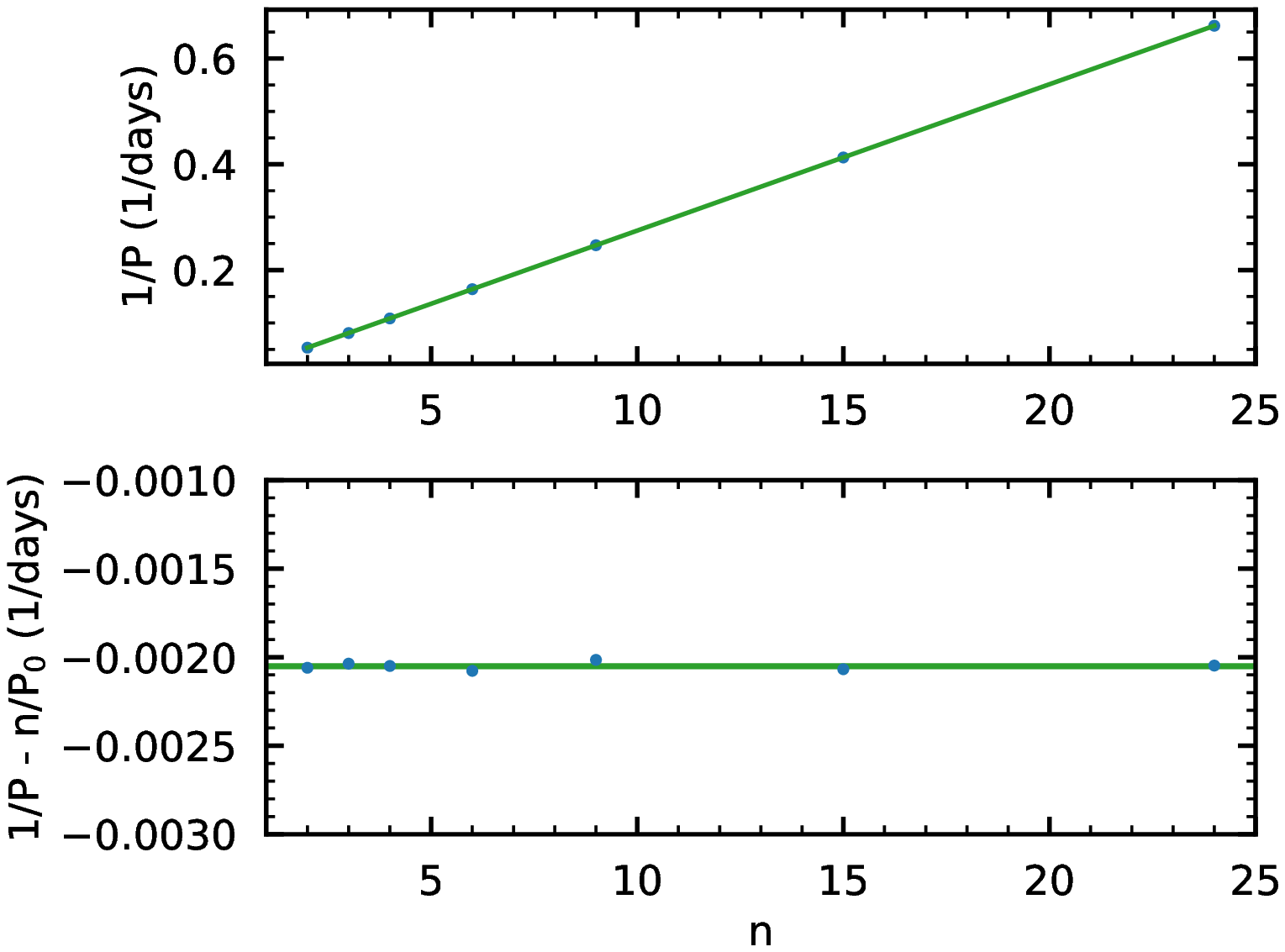}
    \caption{\textit{Top panel}: The blue points (planets h to b, left to right) are the observed orbital frequencies $P^{-1}$ (where the $P$'s are the planets' orbital periods, see Table \ref{table:obs}) as a function of the series of integers $n=[2,3,4,6,9,15,24]$.  When applying a linear fit (green line), the slope corresponds to a period $P_0=0.09~\text{yr}$. \textit{Bottom panel}: plot of $P^{-1}-nP_0^{-1}$ as a function of $n$, showing a residual frequency corresponding to a period of $\sim1.3~\text{yr}$.}
    \label{fig:agol_P0}
    \end{center}
\end{figure}

\begin{figure}
    \begin{center}
    \includegraphics[scale=0.4]{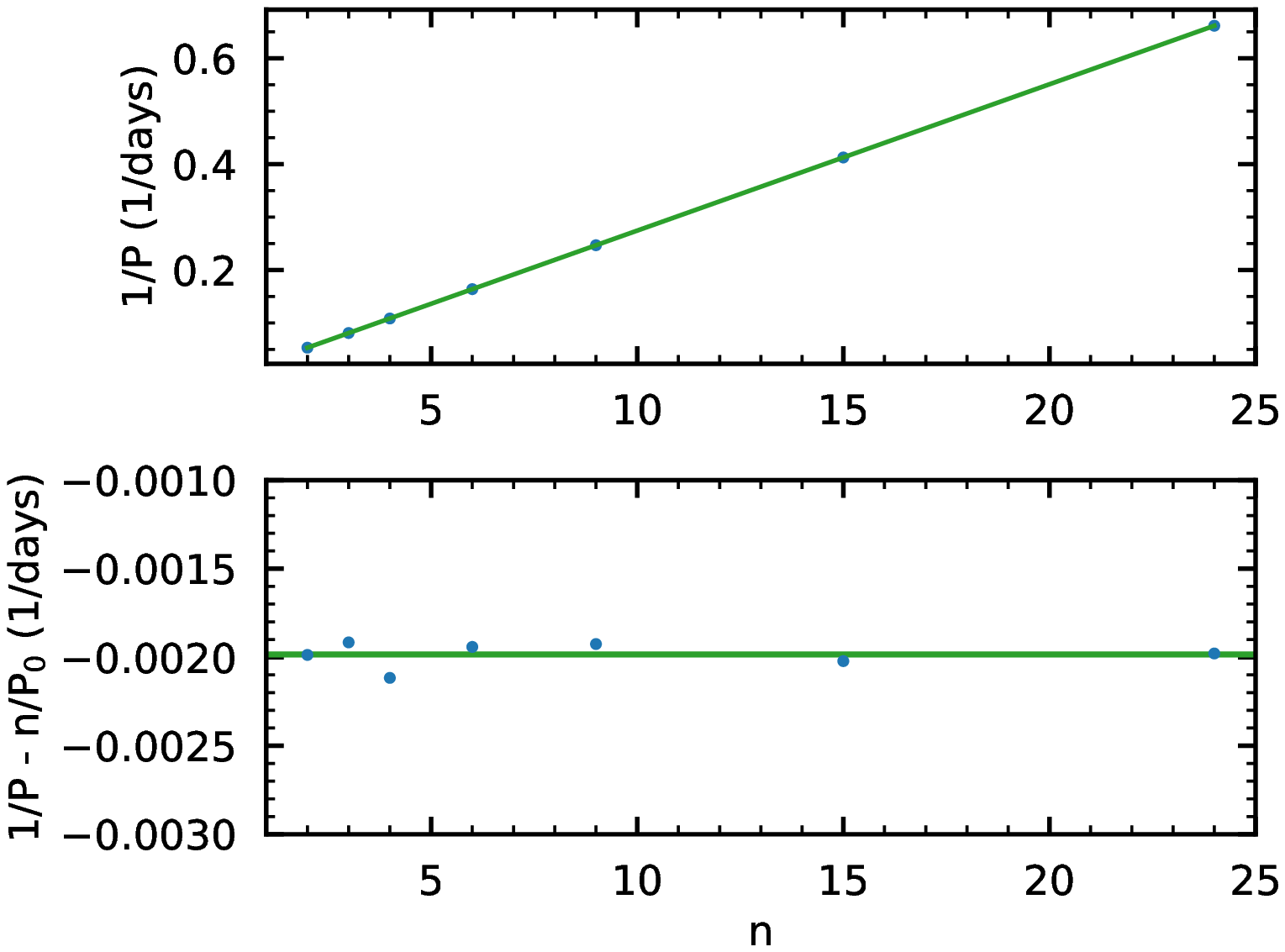}
    \caption{Same as Fig.~\ref{fig:agol_P0}, but with the orbital frequencies obtained via disc migration (see Sect.~\ref{sec:best_disc}).}
    \label{fig:dm_P0}
    \end{center}
\end{figure}

In order to efficiently estimate how close these systems are to the observed one, let us first consider the orbital periods given in Table \ref{table:obs}. The orbital period ratio for each pair of planets suggests that they are close to the ratio between two integers from the following series of integers: $n=[2,3,4,6,9,15,24]$; that is, the orbital frequencies correlate tightly with this integer sequence (the first integer $n=2$ is associated with planet $h$, while $n=24$ is associated with planet $b$). In Fig.~\ref{fig:agol_P0} we plot the planets' orbital frequencies versus this series of integers. As expected, the orbital frequencies nearly lie on the same line, whose slope corresponds to a period of $P_0=36.1$~d, which matches the 0.09~yr period found in our frequency analysis (see Table~\ref{table:ttv_all}). Once this linear variation is subtracted, one can see on the bottom panel that the orbital frequencies are all offset from the period ratios by almost exactly the same frequency, with very small residuals ($\lesssim 10^{-4}~{\rm d}^{-1}$). This frequency of $\sim 0.002~\text{d}^{-1}$ corresponds to a period of $P_{\rm TTV }=492$~d, which matches the main TTV period of 1.3~yr (as seen in Table~\ref{table:ttv_all}). This suggests that in a frame rotating retrograde to the planets' orbits with a period 492 days, the orbital periods are all close to harmonics of a same period of 36.1~d. In comparison, we show in Fig.~\ref{fig:dm_P0} the same analysis for our best-case migration simulation. Using the same sequence of integers, the slope corresponds to a similar period, $P_0=36.2$~d, and an offset period $P_{\rm TTV}=504$~d (with residuals $\lesssim 1.5\times 10^{-4}~{\rm d}^{-1}$), which is also similar to the observed frequency offset.

\begin{figure}
    \begin{center}
    \includegraphics[scale=0.5]{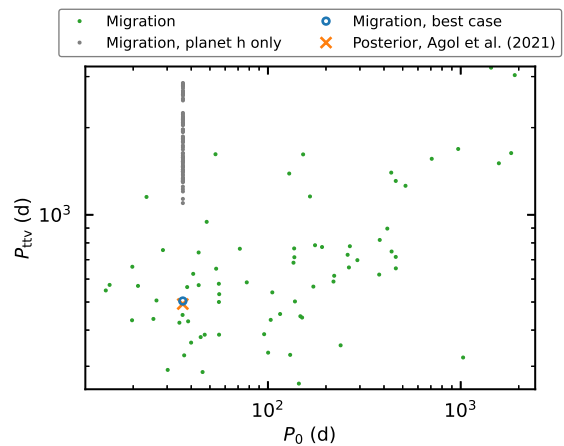}
    \caption{Relevant periods of simulated systems: $P_0$ and $P_{\rm TTV}$ (see Sect.~\ref{sec:discussion}). Green points are for the simulations presented in Sect.~\ref{sec:results} and grey points for the simulations presented in Sect.~\ref{sec:discussion}. The blue circle represents our best-case disc migration simulation (Sect.~\ref{sec:best_disc}), and the orange cross represents the best-fit solution of \citet{agol21}. }
    \label{fig:P0Pttv}
    \end{center}
\end{figure}

The two parameters, $P_0$ and $P_{\rm TTV}$, allow us to estimate whether or not simulations are close to the observed system. For the disc migration simulations presented in Sect.~\ref{sec:results}, as well as the simplified simulations presented in this section, we fitted the same series of integers to the final periods, and deduced the values of $P_0$ and $P_{\rm TTV}$. We discarded systems with  a scatter in the residuals which is larger than $2\times 10^{-4}~{\rm d}^{-1}$. The results are shown in Fig.~\ref{fig:P0Pttv}. The simulations presented in this section (where only planet h is migrating) all have a similar $P_0$ since they started very close to their observed period ratios, but present a large scatter in $P_{\rm TTV}$, and none of them match the observed $P_{\rm TTV}$. Among the simulations presented in Sect.~\ref{sec:results}, the particular simulation that we illustrated in Sect.~\ref{sec:best_disc} (blue circle in Fig.~\ref{fig:P0Pttv}) is the closest to the observed system (orange cross). This method can quickly classify systems depending on whether their period ratios are close to a resonant chain, and identify those simulated systems which resemble the observed system.

\section{Conclusion}
\label{sec:conclusion}

In this work we have studied the dynamics and formation of the best-fit solution to the TRAPPIST-1 system given by \citet{agol21}. In the first part, we focused on the dynamics acting on timescales accessible by future observations. In particular we focus on a short timescale of 4~yr corresponding to the extent of the existing dataset published by \citet{agol21}, and on a longer timescale of 100~yr which reveals dynamical effects visible on this time frame. Our frequency analysis has revealed an intricate chain of resonant interactions that shapes the dynamics of the system. We find that the dynamics is dominated by two-body resonances on short timescales, and by three-body resonances on long timescales. This is in agreement with the analysis of \citet{libert13} which shows that three-body resonances impact the TTV signal on timescales much longer than two-body resonances. We show that all the planets are simultaneously in two-body and three-body resonances, except for the innermost pair. 

In the second part, we carried out $\sim$1000 $N$-body simulations of planets migrating in a disc in an attempt to reproduce the resonant chain of the TRAPPIST-1 system. For planets migrating in discs with a steep inner edge and a strong eccentricity damping ($Q_{\rm e}$ of 0.02--0.1), we found that the outer system was well reproduced by our migration scenario, while the dynamics of the inner system (the eccentricity vectors of planets b and c in particular) did not match that of the best fit solution. Most of the two-body and three-body resonances between the TRAPPIST-1 planets are easily formed, apart from the 8:5 MMR between planets b and c. With regard to these two planets, we found that despite their proximity to the 8:5 MMR, it is the angles associated with the 3:2 MMR which are librating. \citet{papaloizou18} argued that this commensurability could be obtained through tidal expansion of the inner pair away from the 3:2 MMR. Since we do not include tidal interactions with the central star, we chose to focus on the few simulations in which disc migration lead planets b and c to stop near their observed period ratio of 1.6. We presented a particular simulation in which the system at the end of the disc phase was similar to the observed one. We also computed the TTVs at the end of this simulation and showed that they are similar to the ones given by the best-fit solution of \citet{agol21}. This exercise shows that it is possible that the resonant chain of TRAPPIST-1 formed through smooth disc migration. However, owing for the difficulty of locking the innermost pair b-c near its observed period ratio, it appears to be a low-probability event with our choice of initial conditions. Whether tidal effects would change the long-term dynamics of this pair is beyond the scope of this paper, but we note that tidal interactions would damp the free eccentricities, further reducing the amplitude of the corresponding TTVs \citep{lw12}.

In addition to being relevant to future observations, our dynamical analysis puts constraints on the physical processes that took place during the formation and migration of TRAPPIST-1. The TTVs collected these last four years of observation hold key information about the two-body resonant dynamics of TRAPPIST-1. In the coming years, further monitoring of TRAPPIST-1 will supply extended TTVs in which we predict the signature of three-body resonances will become visible. In particular, enough data could be collected soon to recover the periods of 3.3 and 5.1~yr, as well as to give more precise information on the peculiar dynamics of the inner pair of TRAPPIST-1, which may prove to be insightful for formation scenarios.

\section*{Acknowledgements}
The work of JT is supported by a Fonds de la Recherche Scientifique – FNRS Postdoctoral Research Fellowship. The work of ASL is supported by the Fonds de la Recherche Scientifique - FNRS under Grant No. F.4523.20 (DYNAMITE MIS-project). EA acknowledges support from NASA’s NExSS Virtual Planetary Laboratory, funded under NASA Astrobiology Institute Cooperative Agreement Number NNA13AA93A, and the NASA Astrobiology Program grant 80NSSC18K0829. Computational resources have been provided by the Consortium des \'Equipements de Calcul Intensif (C\'ECI), supported by the FNRS-FRFC, the Walloon Region, and
the University of Namur (Conventions No. 2.5020.11, GEQ U.G006.15, 1610468 and RW/GEQ2016).
The research done in this project made use of the SciPy stack \citep{scipy}, including NumPy \citep{numpy} and Matplotlib \citep{matplotlib}, as well as Astropy,\footnote{\url{http://www.astropy.org}} a community-developed core Python package for Astronomy \citep{astropy:2013, astropy:2018}. Simulations in this paper made use of the REBOUND code which is freely available at \url{http://github.com/hannorein/rebound}.

\bibliographystyle{aa}
\bibliography{biblio}

\label{lastpage}

\end{document}